\documentclass[aps,prl,twocolumn,groupedaddress,floats,showpacs]{revtex4}

\usepackage{latexsym}
\usepackage{dcolumn}
\usepackage[dvips]{graphicx}
\usepackage{amssymb}
\usepackage{graphicx}
\usepackage{psfrag, color}
\usepackage{verbatim}
\usepackage{amsmath}
\usepackage{epsf}

\begin{document}

\draft

\title{Dielectric function, screening, and plasmons in 2D graphene}
\author{E. H. Hwang and S. Das Sarma}
\address{Condensed Matter Theory Center, 
Department of Physics, University of Maryland, College Park,
Maryland  20742-4111 } 
\date{\today}

\begin{abstract}
The dynamical dielectric function of two dimensional graphene at 
arbitrary wave vector $q$ and frequency $\omega$, $\epsilon(q,\omega)$,
is calculated in the self-consistent field approximation. 
The results are used to find the dispersion of the plasmon mode and
the electrostatic screening of the Coulomb interaction
in 2D graphene layer within the random phase approximation.
At long wavelengths ($q\rightarrow 0$) the plasmon
dispersion shows the local classical behavior $\omega_{cl}
= \omega_0 \sqrt{q}$, but the density 
dependence of the plasma frequency ($\omega_0 \propto n^{1/4}$) is
different from the usual 2D 
electron system ($\omega_0 \propto n^{1/2}$). 
The wave vector dependent plasmon dispersion and the static screening
function show very different behavior than the usual 2D case.
We show that the intrinsic interband contributions to static graphene
screening can be effectively absorbed in a background dielectric
constant.
\pacs{71.10.-w; 73.21.-b; 73.43.Lp}

\end{abstract}
\vspace{0.5cm}

\maketitle

\section{introduction}

There has been a great deal of recent interest in the electronic
properties of two dimensional (2D)
graphene, a single-layer graphite sheet, both theoretically and
experimentally \cite{Geim,Kim}. The main difference of 2D graphene
compared with 
other (mostly semiconductor-based) 2D materials is the electronic energy
dispersion. In conventional 2D systems, the electron energy with an
effective mass $m^*$ depends quadratically on the momentum, but 
in  graphene, the dispersions of electron and hole
bands are  linear near K, K' points of the Brillouin zone.
Because of the different energy band dispersion,
screening properties in graphene exhibit significantly
different behavior from the conventional 2D systems \cite{rmp}. 
The screening of Coulomb interaction induced by many body effects is 
one of the most important fundamental quantities for understanding
many physical properties. 
For example, dynamical screening determines the elementary excitation
spectra and the collective modes of the electron system, and static
screening determines transport properties through screened Coulomb
carrier scattering by charged impurities. In this paper, we
theoretically obtain the (dynamical and static) screening behavior of
2D graphene by calculating, for the first time, the polarizability and
the dielectric function within the self-consistent field approximation
(i.e. random-phase-approximation (RPA)) for gated-2D graphene free
carrier systems. We apply our 
theory to calculate the 2D graphene plasmon dispersion and the static
screening function, finding some interesting qualitative differences
between graphene and the extensively studied 2D electron systems based on
semiconductor heterostructures and MOSFETs.

In this paper, we calculate the dielectric function of  graphene at
arbitrary wave vector $q$ and frequency $\omega$, $\epsilon(q,\omega)$,
within RPA, in 
which each electron is assumed to move in the self-consistent field
arising from the external field plus the
induced field of all electrons. This is the model which leads to the
famous Lindhard dielectric function for a three-dimensional (3D)
\cite{Lindhard1} and 2D \cite{Lindhard2} 
electron gas. One of the immediate theoretical consequences of the
dielectric function is that its zeros give  the wave vector dependent
plasmon 
mode, $\omega_{pl}(q)$, which is a fundamental elementary 
excitation and a collective density oscillation mode. Using the
theoretical dielectric function we provide 
the plasmon mode dispersion 
both for  single-layer  and 
bilayer graphene. Another important consequence of the dielectric
function is the static screening function which can be obtained 
as the static limit $\omega \rightarrow 0$ of the dielectric
function, describing the electrostatic screening of the electron-electron,
electron-lattice, and electron-impurity interactions.

\section{Polarizability: $\Pi^+$ and $\Pi^-$}

The electron dynamics in 2D graphene is modeled by
a chiral Dirac equation, which describes a linear relation
between energy and momentum.
The corresponding kinetic energy of graphene for 2D wave vector {\bf
  k} is given by
(we use $\hbar = 1$ throughout this paper)
\begin{equation}
\epsilon_{s{\bf k}} = s \gamma |{\bf k}|,
\end{equation}
where $s=\pm 1$ indicate the conduction (+1) and valence ($-1$) bands,
respectively, and $\gamma$ is a band parameter (essentially the 2D
Fermi velocity, which is a constant for graphene instead of being
density dependent).
The corresponding density of states (DOS) is given by
$ D(\epsilon) = g_s g_v |\epsilon|/(2\pi\gamma^2)$, where
$g_s=2$, $g_v=2$ are the spin and valley degeneracies, respectively.
The Fermi momentum ($k_F$) and the Fermi energy ($E_F$) 
of 2D graphene are
given by $k_F = (4\pi n/g_s g_v)^{1/2}$ and $E_F = \gamma k_F$ where
$n$ is the 2D carrier (electron or hole) density.
For the sake of completeness, we also mention that the dimensionless
Wigner-Seitz radius ($r_s$), which measures the ratio of the potential
to the kinetic energy in an interacting quantum Coulomb system
\cite{Lindhard1}, is 
given in doped 2D graphene by $r_s = (e^2/\kappa
\gamma)(4/g_sg_v)^{1/2}$ where $\kappa$ is the background lattice
dielectric constant of the system. We note in the passing the curious
fact that the dimensionless $r_s$ parameter is a {\it constant} in
graphene unlike in the usual 2D ($r_s \sim n^{-1/2}$) and 3D ($r_s
\sim n^{-1/3}$) electron liquids, where $r_s$ (and therefore
interaction effects) increases with decreasing carrier density. The
constancy of $r_s$ in graphene arises trivially from the relativistic
Dirac-like nature of the free carrier graphene dynamics implying that
the `relativistic' effective mass, $m_c = E_F/\gamma^2$, depends on
carrier density precisely as $\sqrt{n}$ cancelling out the
corresponding $\sqrt{n}$ term in the potential energy. Equivalently,
$r_s$ here is just the ``effective fine structure constant'' for
graphene, with a value of $r_s \sim 0.5$ assuming $g_s=g_v=2$ and
$\kappa =4$ (using SiO$_2$ as the substrate material). This small (and
constant) value of graphene $r_s$ indicates it to be a weakly
interacting system for all carrier densities, making RPA an excellent
approximation in graphene since RPA is asymptotically exact in the $r_s
\ll 1$ limit.

In the RPA, the dynamical screening
function (dielectric function) becomes 
\begin{equation}
\varepsilon(q,\omega) = 1+v_c(q) \Pi(q,\omega), 
\end{equation}
where $v_c(q) = 2\pi e^2/\kappa q$ is the 2D
Coulomb interaction,
and $\Pi(q,\omega)$, the 2D
polarizability, is given by the bare bubble diagram
\begin{equation}
\Pi(q,\omega) =-\frac{g_s g_v}{L^2}
\sum_{{\bf k}ss'}\frac{f_{s{\bf k}}-f_{s'{\bf k}'}}
{\omega + \epsilon_{s{\bf k}}-\epsilon_{s'{\bf k}'} +i\eta}F_{ss'}({\bf
  k},{\bf k}'),
\label{pol}
\end{equation}
where
${\bf k}'={\bf k}+{\bf
  q}$, $s,s'=\pm 1$ denote the band indices
and $F_{ss'}({\bf k},{\bf k}')$ is
the overlap of states and given by
$F_{ss'}({\bf k},{\bf k}') = (1 + ss' \cos\theta)/2$, where $\theta$ is the
angle between ${\bf k}$ and ${\bf k}'$, and
$f_{sk}$ is the Fermi distribution function, 
$f_{s{\bf k}} = [\exp \{\beta(\epsilon_{s{\bf k}}-\mu)\} + 1]^{-1}$,
with $\beta = 1/k_BT$ and  $\mu$ the chemical potential.
After performing the summation over $ss'$ we can rewrite the
polarizability as 
\begin{equation}
\Pi(q,\omega) = \Pi^+(q,\omega) + \Pi^-(q,\omega),
\end{equation}
where
\begin{eqnarray}
& &\Pi^+(q,\omega)  = - \frac {g_s g_v}{2L^2}\sum_k \left [
  \frac{[f_{{\bf k}+}-f_{{\bf k}'+}](1+\cos\theta_{kk'})} 
{\omega + \epsilon_{{\bf k}+}-\epsilon_{{\bf k}'+} +i\eta} \right . \nonumber \\
 & + & \left . \frac{f_{{\bf k}+}(1-\cos\theta_{kk'})} 
{\omega + \epsilon_{{\bf k}+} - \epsilon_{{\bf k}'-} +i\eta} 
 - \frac{f_{{\bf
      k}'+}(1-\cos\theta_{kk'})} 
{\omega + \epsilon_{{\bf k}-} - \epsilon_{{\bf k}'+} +i\eta} \right ],
\end{eqnarray}
and
\begin{eqnarray}
& &\Pi^-(q,\omega)  =  -\frac{g_s g_v}{2L^2}\sum_k \left [
  \frac{[f_{{\bf k}-}-f_{{\bf k}'-}](1+\cos\theta_{kk'})} 
{\omega + \epsilon_{{\bf k}-} - \epsilon_{{\bf k}'-} +i\eta} \right . \nonumber \\
& + & \left . \frac{f_{{\bf k}-}(1-\cos\theta_{kk'})} 
{\omega + \epsilon_{{\bf k}-} - \epsilon_{{\bf k}'+} +i\eta} 
 - \frac{f_{{\bf k}'-}(1-\cos\theta_{kk'})} 
{\omega + \epsilon_{{\bf k}+} - \epsilon_{{\bf k}'-} +i\eta} 
\right ]. 
\end{eqnarray}
For intrinsic (i.e. undoped or ungated with $n$ and $E_F$ both being
zero) graphene, in which the conduction band is empty and
the valence band is fully occupied at zero temperature 
(i.e. $E_F=0$), we have $f_{{\bf k}+}=0$ and $f_{{\bf k}-}=1$. Then
the polarizability becomes $\Pi(q,\omega) = \Pi^-(q,\omega)$, which
has been previously obtained in the renormalization group approach
\cite{Guinea}. Recently $\Pi^-(q,\omega)$ has been reconsidered to
discuss screening effects of Coulomb 
interaction in intrinsic graphene \cite{khve}. In general,
$\Pi^{+}(q,\omega)$ does not vanish 
for most systems
because the Fermi energy is typically located in the conduction or
the valence band. 
But graphene is a most peculiar zero-gap semiconductor system where
$E_F=0$ in the intrinsic undoped situation. In the doped or gated
situation $n$, $E_F \neq 0$ in graphene, and now $\Pi^+$ is finite.
In the following we provide the zero temperature
polarizability in the doped or gated case where the Fermi energy is
not zero. 

By introducing the dimensionless quantities $x=q/k_F$ and $\nu =
\omega/E_F$, and $\tilde{\Pi}(q,\omega) = \Pi(q,\omega)/D_0$ where
$D_0 \equiv D(E_F)= (g_sg_vn/\pi)^{1/2}/\gamma$ is the DOS at Fermi
energy, we have 
\begin{equation}
\tilde{\Pi}^+(x,\nu) = \tilde{\Pi}^+_1(x,\nu)\theta(\nu-x) +
\tilde{\Pi}^+_2(x,\nu)\theta(x-\nu),  
\label{pol0}
\end{equation}
where the real parts of the polarizability are given
\begin{eqnarray}
{\rm{Re}}\tilde{\Pi}^+_1(x,\nu) & = &1  -  \frac{1}{8\sqrt{\nu^2-x^2}}
\left \{ f_1(x,\nu)   \theta(|2+\nu|-x) \right . \nonumber \\
& + & {\rm{sgn}} (\nu-2+x)f_1(x,-\nu)
  \theta(|2-\nu|-x) \nonumber \\
& + &
\left .  f_2(x,\nu)[\theta(x+2-\nu)+\theta(2-x-\nu)] \right \}
\end{eqnarray}
\begin{eqnarray}
{\rm{Re}}\tilde{\Pi}^+_2(x,\nu)  =1 - \frac{1}{8\sqrt{x^2-\nu^2}}
\left \{ f_3(x,\nu)\theta(x-|\nu+2|) \right . \nonumber \\
 +  f_3(x,-\nu)\theta(x-|\nu-2|) \nonumber \\
+ \left . \frac{\pi x^2}{2}[\theta(|\nu+2|-x) + \theta(|\nu-2|-x)]
\right \}
\label{pip2}
\end{eqnarray}
and the imaginary parts of the polarizability are
\begin{eqnarray}
{\rm{Im}} \tilde{\Pi}^+_1(x,\nu)& =& \frac{-1}{8\sqrt{\nu^2-x^2}} [f_3(x,-\nu)
  \theta(x-|\nu-2|) \nonumber \\
& +& \frac{\pi x^2}{2} [\theta(x+2-\nu) + \theta(2-x-\nu)]]
\end{eqnarray}
\begin{eqnarray}
{\rm Im} \tilde{\Pi}^+_2(x,\nu)  &=& 
\frac{\theta(\nu-x+2)}{8\sqrt{x^2-\nu^2}}
[f_4(x,\nu)\nonumber \\
& -& f_4(x,-\nu)\theta(2-x-\nu)],
\end{eqnarray}
where
\begin{eqnarray}
f_1(x,\nu) & = & (2+\nu)\sqrt{(2+\nu)^2-x^2} \nonumber \\
& - &  x^2 \ln
\frac{\sqrt{(2+\nu)^2-x^2} + (2+\nu)}{|\sqrt{\nu^2-x^2} + \nu|}
\end{eqnarray}
\begin{equation}
f_2(x,\nu) = x^2 \ln \frac{\nu-\sqrt{\nu^2-x^2}}{x}
\end{equation}
\begin{equation}
f_3(x,\nu) = (2+\nu)\sqrt{x^2-(2+\nu)^2} + x^2
\sin^{-1}\frac{2+\nu}{x} 
\end{equation}
\begin{eqnarray}
f_4(x,\nu)& = &(2+\nu)\sqrt{(2+\nu)^2-x^2} \nonumber \\
     & - & x^2\ln
\frac{\sqrt{(2+\nu)^2-x^2} +(2+\nu)}{x}
\end{eqnarray}
and $\tilde{\Pi}^{-}(x,\nu)$ can be calculated to be
\begin{equation}
\tilde{\Pi}^{-}(x,\nu) = \frac{\pi x^2
  \theta(x-\nu)}{8\sqrt{x^2-\nu^2}} + i \frac{\pi x^2
  \theta(\nu-x)}{8 \sqrt{\nu^2-x^2}}.
\label{pim}
\end{equation}
Eqs. (\ref{pol0})-(\ref{pim}) are the basic results obtained in this
paper, giving the 2D doped graphene polarizability analytically. Note
that our 2D graphene polarizability is completely different from the
corresponding 2D Lindhard function first calculated in
ref. \onlinecite{Lindhard2}, which is appropriate for the usual 2D
systems with parabolic band dispersion.

\section{plasmons in RPA}

As a significant consequence of the dielectric function we calculate
the long wavelength plasmon dispersion for single-layer graphene and
for bilayer graphene.
The longitudinal collective-mode dispersion, or plasmon mode dispersion, 
can be calculated by looking for poles of the
density correlation function, or equivalently, by looking for zeros of the
dynamical dielectric function, $\epsilon({\bf q},\omega) = 1 -
v(q)\Pi({\bf q},\omega)$.
In the long wavelength limit ($q \rightarrow 0$) we have the
following limiting forms in the high- and low-frequency regimes:
\begin{eqnarray}
\Pi(q,\omega)  \approx \left \{
\begin{array}{ll}  
\frac{D_0\gamma^2q^2}{2\omega^2} \left [ 1-\frac{\omega^2}{4E_F^2}
\right ],  & \gamma q < \omega < 2E_F  \\
D_0 \left [ 1 + i \frac{ \omega}{\gamma q} \right ], & \omega <
\gamma q. 
\end{array} 
\right .
\label{approx}
\end{eqnarray}
In the $q \rightarrow 0$ limit, we have the plasmon mode
dispersion $\omega_p(q)$ for a single-layer graphene as
\begin{equation}
\omega_{cl} \equiv \omega_{p}(q \rightarrow 0) = \omega_0 \sqrt{q}
\end{equation}
where $\omega_0 = (g_s g_v e^2 E_F/2\kappa)^{1/2}$. 
The leading order (or 
local) plasmon has exactly the same dispersion, $q^{1/2}$, as the
normal 2D plasmon \cite{rmp}. 
However, the density dependence of the plasma frequency in
graphene shows a different behavior, i.e. $\omega_0 \propto n^{1/4}$
compared with the classical 2D plasmon behavior
where $\omega_0 \propto n^{1/2}$. This is
a direct  consequence of the quantum relativistic nature of graphene.
Even though the long wavelength 
plasmons have identical dispersions for both cases, 
the dispersion calculated within RPA
including finite-wave-vector nonlocal, (i.e. higher order in q)
effects show very different behavior. In normal 2D
\cite{rmp,Lindhard2} the non-local
correction leads to an increase in plasma
frequency, [$\omega_p(q)/\omega_{cl} = 1 + (3/4)(q/q_{TF})$], where
$q_{TF}=g_sg_vme^2/\kappa$ is the usual 2D Thomas-Fermi wave vector),  
but in graphene the correction within RPA leads to a
decrease in plasma frequency compared with $\omega_{cl}$ 
[$\omega_p(q)/\omega_{cl} = 1 - q_0q/8k_F^2$], where $q_0 =
g_sg_ve^2k_F/\gamma\kappa$ is the corresponding graphene Thomas-Fermi wave
vector. Recently, the plasmon mode of graphene 
has been considered numerically in the presence of spin-orbit coupling
\cite{wang}.

For bilayer graphene, {\it without inter-layer hopping}, we have the
leading order $q$ dependence 
of the collective modes by solving a two component determinantal
equation \cite{dassarma}
\begin{eqnarray}
\omega_+(q) &\approx &  \omega_0 \sqrt{2q} \nonumber \\
\omega_-(q) &\approx & 2 \omega_0 \sqrt{d} q,
\end{eqnarray}
where $d$ is the layer separation between the two 2D graphene sheets. 
The $\omega_+$ mode, the
optical-plasmon mode (in-phase mode of the coupled system), has the
well known $q^{1/2}$ behavior,
independent of the layer separation $d$ at long wavelengths.
The other mode $\omega_-$ is the
acoustic plasmon mode (out-of-phase mode of the coupled system) which goes
as $q$ in long wavelengths and depends on the separation $d$. Thus,
the coupled plasmons in graphene 
show the same long wavelength behaviors as those of normal 2D systems. 
But, the density dependences of the plasma frequency and the large wave
vector dependences are again very different from the corresponding
normal 2D systems \cite{dassarma}.
When interlayer hopping is included in a bilayer system, the
in-phase plasmon mode is 
qualitatively unaffected by tunneling. However, the out-of-phase
plasmon mode develops a long wavelength gap (depolarization shift) in
the presence of tunneling \cite{hwang}, i.e. $\omega_-(q \rightarrow
0) = (2t_{\perp})^2(1 + q_0d)$, where $t_{\perp}$ is the inter-layer hopping. 
Due to strong interlayer coupling in bilayer graphene, in general, we
have $\omega_+ \ll \omega_-$ at long wavelengths.

\begin{figure}
\epsfysize=1.7in
\centerline{\epsffile{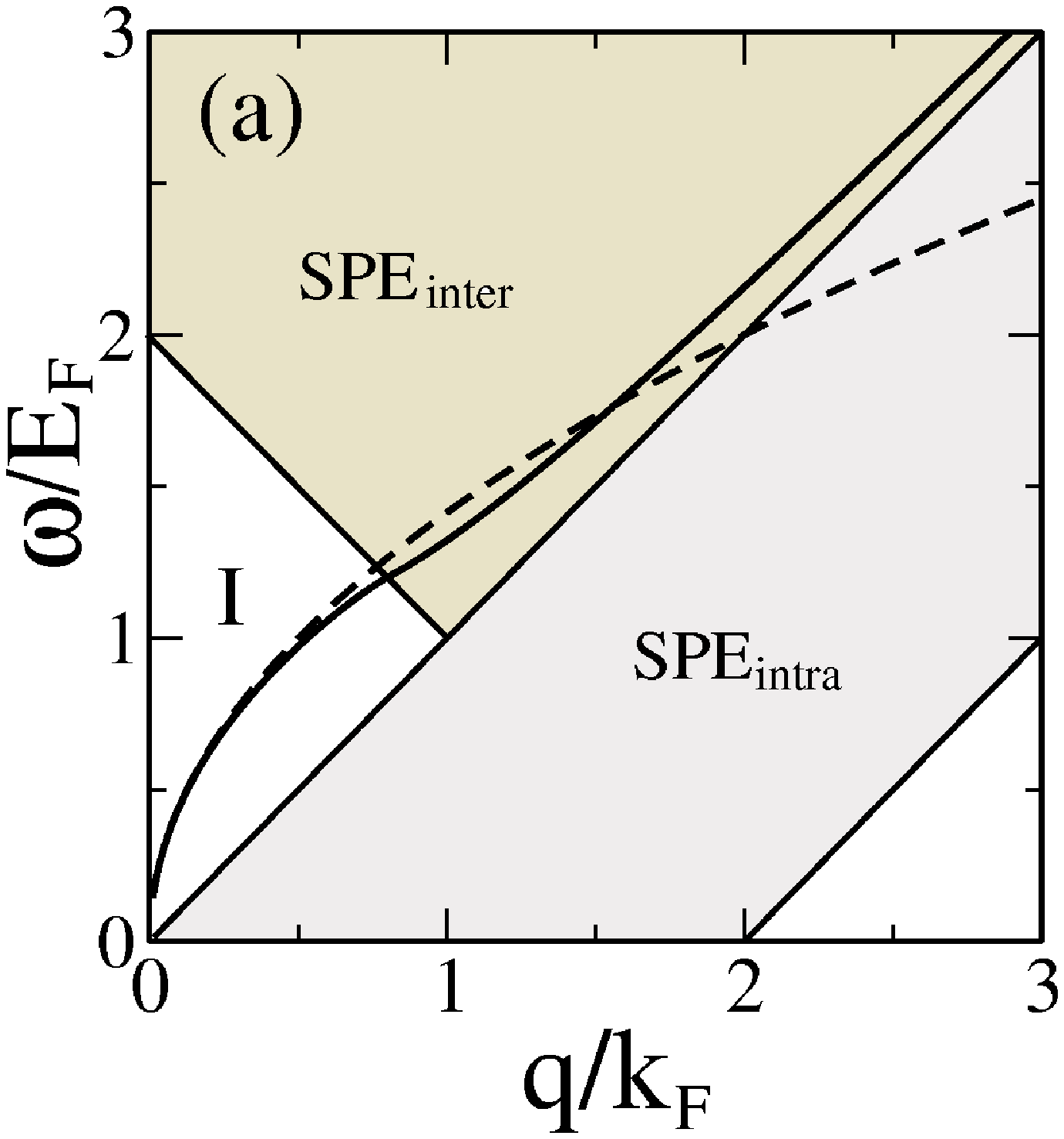}}
\centerline{\epsffile{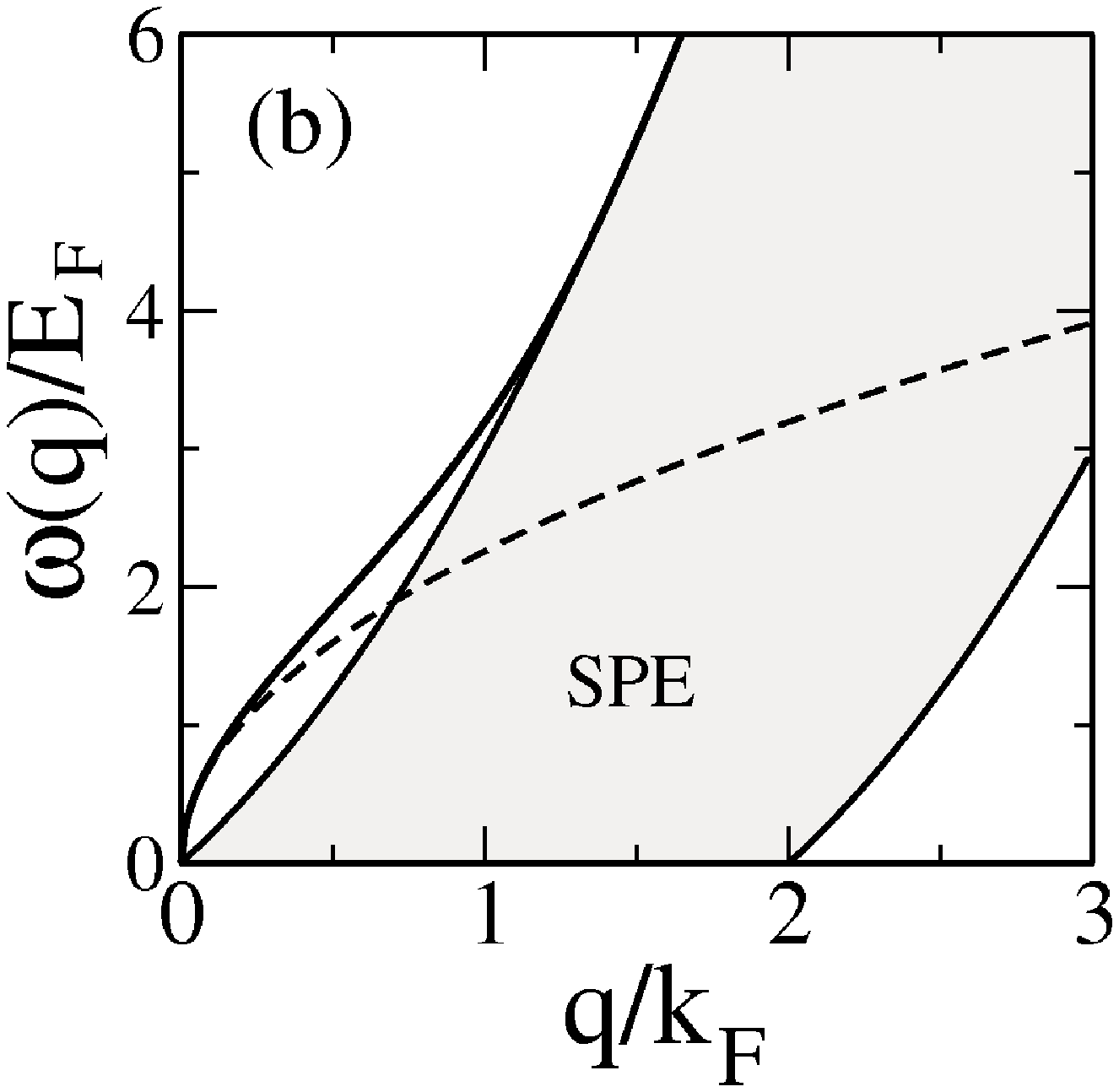}}
\caption{
(a) Plasmon mode dispersion in 2D graphene (solid thick
line) calculated within RPA. The dashed line indicates the local
long wavelength plasmon 
dispersion. Thin solid lines represent the boundaries of the
single particle excitation (SPE) Landau damping regime for intra- and
inter-band electron-hole excitations.
(b) The plasmon dispersion and SPE for a normal 2D system with a
quadratic energy dispersion.
}
\label{plasmon}
\end{figure}

In Fig. \ref{plasmon} we show the calculated plasmon dispersion
within RPA (solid line) compared with the classical local plasmon (dashed
line). We use the following parameters: $\kappa = 2.5$, $\gamma =
6.5$eV\AA, and a density $n=10^{12}$ cm$^{-2}$. 
In fig. \ref{plasmon}(b) we show the corresponding 2D regular plasmons
with n-GaAs
parameters ($n=5 \times 10^{11}$ cm$^{-2}$).
In Fig. \ref{plasmon} we also show  
the electron-hole continuum or single particle
excitation (SPE)  region in ($q, \omega$) space, which  determines the
absorption (Landau damping) of the external 
field at given frequency and wave vector.
The SPE continuum is defined by the non-zero
value of the imaginary part of the polarizability function,
Im$\Pi(q,\omega) \neq 0$. 
For a normal 2D system only indirect ($q\neq 0$) transition is
possible within 
the band, and the SPE boundaries are given by $\omega_{1,2} = q^2/2m
\pm qk_F/m$. 
However, for 2D graphene both intraband and interband transition are
possible, and the boundaries are given in Fig. 1(a).
The intraband SPE boundaries are $\omega_1
=  \gamma q$ (upper boundary) and $\omega_2 = 0$ for $q<2k_F$, $\omega_2 =
\gamma q - 2E_F$ for $q>2k_F$ (lower boundary). 
The direct transition
($q=0$) is also possible from the valence band to the empty conduction
band. Due to the phase-space restriction 
the interband SPE continuum has a 
gap at small wave vectors. For $q=0$, the transition
is not allowed at $0 < \omega < 2E_F$. 
If the collective mode lies inside the SPE
continuum we expect the mode to be damped. Since the normal 2D plasmon
lies, at long wave lengths, above the SPE continuum it never decays to
electron-hole 
pair within RPA. But for graphene the plasmon lies inside the
interband SPE continuum decaying into electron-hole pairs. Only in the region
I of Fig. 1(a) the plasmon is not damped.
The other different feature between a normal 2D plasmon and a
graphene plasmon occurs at large wave vectors.
The normal 2D plasmon mode enters into the
SPE continuum at a critical wave vector, and therefore does not exist at very
high wave vectors. All spectral weight
of the plasmon mode is transferred to the SPE.
But the graphene plasmon does not enter into the intraband SPE and exists
for all wave vectors, except for its decay into real interband
electron-hole pairs in the SPE$_{\rm inter}$ regime.

\begin{figure}
\epsfysize=2.2in
\centerline{\epsffile{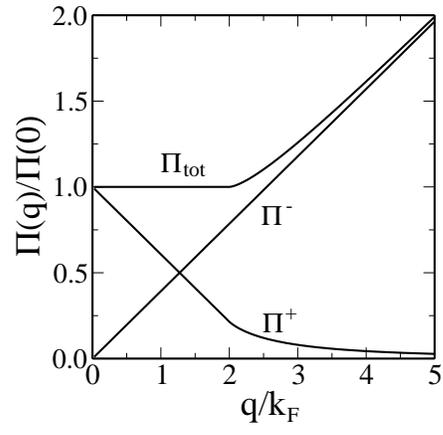}}
\caption{
Static polarizability for 2D graphene:
$\Pi_{tot} = \Pi^{+} + \Pi^-$.
Here
$\Pi(0) = D(E_F) = g_s g_v k_F/2 \pi \gamma$. 
}
\label{s_pol}
\end{figure}

\section{static screening}

Now we consider the static polarizability $\Pi(q,\omega=0)$.
From Eq. (\ref{pip2}) we have
\begin{equation}
\tilde{\Pi}^+(q)  = \left \{
\begin{matrix}
1 - \frac{\pi q}{8k_F}, & q \le 2k_F \cr
1-\frac{1}{2}\sqrt{1-\frac{4k_F^2}{q^2}} -
\frac{q}{4k_F}\sin^{-1}\frac{2k_F}{q} , &
q > 2k_F  \cr
\end{matrix} 
\right .
\label{spi}
\end{equation}
and from Eq. (\ref{pim}) we have 
\begin{equation}
\tilde{\Pi}^-(q) = {\pi  q}/{8k_F}. 
\label{spim}
\end{equation}
Thus, the total static polarizability
becomes a constant at $q \le 2 k_F$ as in a normal 2D systems,
i.e. $\Pi(q) =\Pi^+(q) + \Pi^-(q) = D(E_F)$ for $q \le 2k_F$. 
In Fig. \ref{s_pol} we show the calculated static polarizability
as a function of wave vector. 
For a normal 2D system the screening wave vector, $q_s = q_{TF}=g_s
g_v m e^2 /\kappa$, is 
independent of electron 
concentration, but for 2D graphene the screening wave vector is
given by $q_s = g_s g_v e^2 k_F /\kappa \gamma$ which is proportional to
the square root of the density, $n^{1/2}$.
In the large momentum transfer regime, $q > 2k_F$, the static screening
increases linearly with $q$
due to the interband transition. This is a very different behavior
from a normal 2D system where the static polarizability falls
off rapidly for $q >2k_F$ with a cusp at $q=2k_F$ \cite{rmp}. 
The linear increase of the
static polarizability with $q$ gives rise to 
an enhancement of the effective dielectric constant
$\kappa^* (q \rightarrow \infty)= \kappa (1 + g_sg_v\pi r_s/8)$ in
graphene.  Note that in a normal 2D system
$\kappa^* \rightarrow \kappa$ as $q \rightarrow \infty$. Thus, the
effective interaction in 2D
graphene decreases at short wave lengths due to polarization
effects. This large wave vector screening behavior is typical of an
insulator. Thus, 2D graphene screening is a combination of
``metallic'' screening (due to $\Pi^{+}$) and ``insulating''
screening (due to $\Pi^-$), leading to overall rather strange
screening properties, all of which can be traced back to the zero-gap
chiral relativistic nature of graphene.

In may be worthwhile to ask whether the {\it intrinsic} graphene
contribution, arising strictly from the interband transitions due to
the filled valence band (i.e. the $\Pi^-$ term in our graphene
polarizability), can be absorbed in the effective background lattice
dielectric constant $\kappa$ just as one does in a regular
semiconductor ($\kappa_{Si}=11.5$; $\kappa_{GaAs} = 12.9$) or
insulator ($\kappa_{SiO_2} = 3.9$) in discussing free carrier
screening by doping or gating indeed free carriers in conduction
(electrons) or valence (holes) bands. In particular, only {\it
  intraband} free carrier screening is explicitly considered in the
usual 2D screening function \cite{Lindhard2,rmp} extensively used
\cite{stern,hwang} in the quantitative analysis of quantum transport
in 2D semiconductor devices such as Si MOSFETs, GaAs modulation-doped
high-mobility transistors (HEMTs), and undoped gated GaAs
heterostructures (HIGFETs). The interband transition induced screening
in the semiconductor based 2D structures is included in the theory
simply by appropriately modifying the effective background lattice
dielectric constant from the usual vacuum value of unity to a value
around ten.

To see whether the effect of the interband $\Pi^-$ polarizability can
be `trivially' absorbed in a background lattice dielectric constant we
rewrite a 2D graphene dielectric function $\epsilon(q) =
1+ v_c(q)\Pi(q)$ to obtain:
\begin{equation}
\epsilon(q) = 1 + \frac{2\pi e^2}{\kappa q} \left [ \Pi^-(q) +
  \Pi^+(q) \right ].
\end{equation}
Using Eq. (\ref{spim}),
$\Pi^-(q) = D(E_F) {\pi q}/{8 k_F}$,
we get
$v_c(q)\Pi^-(q) = \frac{g_sg_v \pi}{8} \frac{e^2}{\kappa \gamma} =
\frac{g_s g_v \pi}{8} r_s$.
Thus, we have
\begin{equation}
\epsilon(q) = 1 + \frac{g_sg_v \pi}{8}r_s + v_c(q) \Pi^{+}(q).
\end{equation}
Introducing an effective intrinsic background graphene dielectric
constant 
\begin{equation}
\kappa^* = 1+g_sg_v \pi r_s/8, 
\end{equation}
we have
\begin{equation}
\epsilon(q) = \kappa^*\left [ 1 + \frac{2\pi e^2}{\kappa \kappa^* q}
  \Pi^+(q) \right ].
\end{equation}
Writing an effective {\it free carrier} 2D graphene dielectric
function 
\begin{equation}
\epsilon^+(q) \equiv 1 + v_c^+(q) \Pi^+(q),
\end{equation}
where $v_c^+(q) = 2\pi e^2/\kappa \kappa^* q$, we have
\begin{equation}
\epsilon(q) \equiv \kappa^* \epsilon^+(q).
\label{epsstar}
\end{equation}
Eq. (\ref{epsstar}) shows that the intrinsic screening contribution
arising from the interband $\Pi^-$ term can be completely subsumed by
introducing an effective graphene background lattice
dielectric constant $\kappa^* = 1 + g_s g_v \pi r_s/8$, and by making
the replacement $\kappa \rightarrow \kappa \kappa^*$
throughout. Introducing the effective dielectric constant $\kappa^*$
allows one to use only the free carrier screening function
\begin{equation}
\epsilon^+(q) = 1 + \frac{2\pi e^2}{\kappa \kappa^* q} \Pi^+(q),
\end{equation}
for describing free carrier screening properties of 2D graphene. We
note that $\kappa$ itself here is the background lattice dielectric
constant arising from the insulating substrate (i.e. SiO$_2$ in most
situations) with $\kappa = (1+\kappa_{SiO_2})/2 \approx 2.5$, and
$\kappa^* = 1 + g_sg_v \pi r_s/8 \approx 2.3$ with $r_s = e^2/\kappa
\hbar \gamma \approx 0.7$ for graphene on SiO$_2$ substrate. Thus, the
substitution $\kappa \rightarrow \kappa \kappa^*$ arising from the
interband contributions enhances the effective background graphene
dielectric constant to $\kappa \kappa^* \approx 6$ --- this
approximate factor of 2 increase in the background dielectric constant
arising from interband contributions further suppress Coulomb
interaction effects in extrinsic graphene.

The large $q$ behavior of graphene dielectric screening, $\epsilon(q
\rightarrow \infty) \rightarrow \kappa^* = 1+g_sg_v \pi r_s/8$, again
demonstrates that the short-wavelength screened Coulomb potential in
2D graphene goes as $2\pi e^2/\kappa \kappa^*q$, with an enhanced
background effective dielectric constant $\kappa \kappa^*$ where
$\kappa$ is the effective background dielectric constant arising from
the substrate and $\kappa^*$ arising from graphene interband
polarizability $\Pi^-$ as discussed above. This implies that a
suspended 2D graphene film, without any substrate (i.e. $\kappa = 1$),
would have an effective background lattice dielectric constant of
$\kappa^* = 1 + g_sg_v \pi r_s/8$ with $r_s = e^2/\hbar \gamma$ (since
$\kappa = 1$). Putting in $g_sg_v=4$, we get $\kappa^* \approx
4$. Thus the background static lattice dielectric constant of
intrinsic graphene, due to interband transitions, is around 4.

If the $T=0$ transport properties of graphene are dominated by charged
impurity scattering as is thought to be the case \cite{hwang_g} then
the long-wavelength Thomas-Fermi screening becomes an important input
for calculating the screened charged impurity potential:
$\epsilon_{TF}(q) \equiv \epsilon_{RPA}(q\rightarrow 0)$ becomes
\begin{equation}
\epsilon_{TF}(q) = 1 + q_{TF}/q,
\label{TF1}
\end{equation}
where $q_{TF} \equiv q_s = g_sg_ve^2k_F/\kappa\gamma$. Note that one
can equivalently define a long wavelength effective TF screening
function $\epsilon_{TF}^+(q)$ where interband screening effects are
absorbed in an effective background dielectric constant $\kappa^*$;
\begin{equation}
\epsilon^+_{TF}(q) = 1 + q_{TF}^*/q,
\label{TF2}
\end{equation}
where $q_{TF}^* = q_{TF}/\kappa^*$. As discussed above, the identity,
$\epsilon(q) \equiv \kappa^* \epsilon^+(q)$, guarantees the
equivalence between Eqs. (\ref{TF1}) and (\ref{TF2}) for long
wavelength graphene screening. We note also that the $q=0$ screening
wavevector $q_{TF}$ is simply proportional to the graphene density of
states at the Fermi level at $T=0$.

Although the two screening descriptions, based on $\epsilon(q)$ with
the background dielectric constant being just $\kappa$ and on
$\epsilon^+(q)$ with the background dielectric constant being $\kappa
\kappa^*$, are precisely equivalent for $T=0$ static screening
properties, the two descriptions are {\it not} inequivalent at finite
temperatures. Therefore, it is more appropriate to use the full
dielectric function $\epsilon(q)$ in theoretical work on 2D graphene.

\section{conclusion}

In conclusion, we have theoretically obtained analytic expressions for
doped (i.e. $E_F \neq 0$) 2D extrinsic graphene polarizability,
dielectric function, 
plasmon dispersion, and static screening properties, finding a number
of intriguing qualitative differences with the corresponding normal
(and extensively studied) 2D electron systems. The differences, with
interesting observable consequences, can all be understood as arising
from the zero-band gap intrinsic nature of undoped graphene with
chiral linear relativistic bare carrier energy band dispersion. Some
of our qualitatively new predictions, such as the $n^{1/4}$ dependence
of the long-wave length graphene plasma frequency in contrast to the
well-known $n^{1/2}$ behavior of classical and normal 2D plasmons,
should be easily verifiable experimentally using the standard
experimental techniques of infra-red absorption \cite{Tsui} and/or
inelastic light scattering \cite{Olego} spectroscopies. Similarly, our
prediction of the peculiar nature of the graphene plasmon damping
(i.e. {\it no} Landau damping due to intraband electron-hole pairs,
but finite Landau damping due to interband electron-hole pairs) should
be easily verifiable. Our predicted different screening behavior in
graphene at large wave vector should have consequences for transport
properties. Our RPA theory should be an excellent qualitative
approximation for 2D graphene properties at all carrier densities (as
long as the system remains a {\it homogeneous} 2D carrier system,
which may not be true for $ n \alt 10^{12}cm^{-2}$) since the effective
$r_s$-parameter for graphene is a constant ($<1$), making RPA
quantitatively accurate in graphene.
Finally, we point out that the effective Fermi temperature, $T_F =
E_F/k_B$, being very high ($\sim 1300K$ for $n\sim 10^{12}$ cm$^{-2}$)
in graphene, our $T=0$ theory should apply all the way to room
temperatures. 
We note that the long wavelength dielectric function for bulk graphite
was earlier considered 
within an approximation scheme in ref. \onlinecite{Shung}, and the zero
frequency limit was 
recently considered in ref. \onlinecite{Ando}.

Before concluding, we point out that some aspects of graphene
collective modes and linear response have been discussed in the recent
literature. In particular, the intrinsic situation without any free
carriers has been considered in \cite{khve} whereas our emphasis in
this work has been {\it extrinsic} graphene with free carriers
(electrons/holes in conduction/valence band) induced by external
gating or doping. There has been a recent purely numerical study
\cite{wang} of graphene collective mode spectra in the presence of
spin-orbit coupling.

{\it Note added} After submitting the present paper, we became aware
of related work \cite{wunsch}.


This work is supported by the US-ONR, the LPS, and the Microsoft
corporation. 


\end{document}